\documentclass[sigconf,review=false,screen=false]{acmart}
\usepackage{hyperref}
\usepackage{hyperxmp}
\usepackage[utf8]{inputenc}
\usepackage{algorithm}
\usepackage{algpseudocode}
\usepackage{amsmath}
\usepackage{subfigure}
\usepackage{multirow} 
\usepackage{subcaption}

\DeclareUnicodeCharacter{2217}{*}

\copyrightyear{2025}
\acmYear{2025}
\setcopyright{rightsretained}
\acmConference[CHI EA '25]{Extended Abstracts of the CHI Conference on Human Factors in Computing Systems}{April 26-May 1, 2025}{Yokohama, Japan}
\acmBooktitle{Extended Abstracts of the CHI Conference on Human Factors in Computing Systems (CHI EA '25), April 26-May 1, 2025, Yokohama, Japan}\acmDOI{10.1145/3706599.3719995}
\acmISBN{979-8-4007-1395-8/2025/04}

\makeatletter
\gdef\@copyrightpermission{
    This work is licensed under a Creative Commons Attribution International 4.0 License.
    \url{https://creativecommons.org/licenses/by/4.0/}
\vspace{5pt}
}
\makeatother

\author{Niall McGuire}
\email{niall.mcguire@strath.ac.uk}
\affiliation{%
  \institution{NeuraSearch Laboratory, University of Strathclyde}
  \streetaddress{P.O. Box 1212}
  \city{Glasgow}
  \country{UK}
  \postcode{43017-6221}
}

\author{Yashar Moshfeghi}
\email{yashar.moshfeghi@strath.ac.uk}
\affiliation{%
  \institution{NeuraSearch Laboratory, University of Strathclyde}
  \streetaddress{P.O. Box 1212}
  \city{Glasgow}
  \country{UK}
  \postcode{43017-6221}
}

\ccsdesc[300]{Human-centered computing~Collaborative interaction}
\ccsdesc[500]{Human-centered computing~Empirical studies in HCI}

\begin{document}


\title{On Error Classification from Physiological Signals within Airborne Environment}









\begin{abstract}
Human error remains a critical concern in aviation safety, contributing to 70-80\% of accidents despite technological advancements. While physiological measures show promise for error detection in laboratory settings, their effectiveness in dynamic flight environments remains underexplored. Through live flight trials with nine commercial pilots, we investigated whether established error-detection approaches maintain accuracy during actual flight operations. Participants completed standardized multi-tasking scenarios across conditions ranging from laboratory settings to straight-and-level flight and 2G manoeuvres while we collected synchronized physiological data. Our findings demonstrate that EEG-based classification maintains high accuracy (87.83\%) during complex flight manoeuvres, comparable to laboratory performance (89.23\%). Eye-tracking showed moderate performance (82.50\%), while ECG performed near chance level (51.50\%). Classification accuracy remained stable across flight conditions, with minimal degradation during 2G manoeuvres. These results provide the first evidence that physiological error detection can translate effectively to operational aviation environments.
\end{abstract}

\maketitle

\section{Introduction}
Despite decades of technological progress and regulatory improvements in aviation, human error continues to be implicated in 70-80\% of aircraft accidents and incidents \cite{wiegmann2017human}. While current approaches rely on post-incident analysis and self-reporting, these methods cannot enable real-time intervention when errors occur \cite{reason1990human}. Prior advances in wearable physiological sensors offer a promising new direction -- the potential to detect errors as they emerge, enabling proactive intervention \cite{dehais2020neuroergonomics}. Laboratory studies have demonstrated that physiological signals like EEG, ECG and eye movements can indicate error states with high accuracy \cite{gehring1993neural}. However, transitioning these findings from controlled settings to the dynamic aviation environment presents significant challenges. Aircraft motion, electromagnetic interference, and operational stresses could all impact signal quality and reliability \cite{wilson2005eeg}. 

To the best of our knowledge, no prior work has validated error detection from physiological signals in airborne scenarios.  We conducted a novel investigation examining whether established error-detection approaches from laboratory studies can translate effectively to the airborne environment. Through a series of live flight trials with nine participants, we collected synchronized physiological data (EEG, ECG, eye tracking) while participants completed controlled multi-tasking scenarios across different flight conditions. Our preliminary results suggest EEG-based detection achieves 87.8\% accuracy even during complex manoeuvres, comparable to laboratory performance \cite{chavarriaga2010learning}. These findings lay the groundwork for developing adaptive cockpit interfaces that could enhance aviation safety through early error detection and intervention \cite{dehais2017pilot}. This work makes three key contributions to human-computer interaction research: 1) a methodology for eliciting and measuring human error across varying environmental conditions; 2) a framework for assessing the translation of laboratory-derived physiological measurement approaches to dynamic operational settings; and 3) quantitative evidence examining the feasibility of multi-modal physiological monitoring in aviation contexts.

\section{Related Work}
Traditional approaches to error detection in aviation have primarily relied on observer ratings, self-reporting, and post-flight analysis \cite{wiegmann2017human}. While valuable, these methods suffer from significant limitations - they are retrospective, subject to perception bias, and crucially, cannot enable real-time intervention \cite{yeh1988dissociation, li2001factors}. Prior work has explored automated detection systems based on behavioural metrics \cite{wickens2017effects}, but these approaches often fail to capture the cognitive states that precede error commission. This gap between detection and intervention capabilities has motivated the exploration of physiological approaches in human-computer interaction research \cite{fairclough2009fundamentals, dehais2017pilot}. Laboratory studies have identified several promising physiological indicators of error states. EEG research has revealed specific error-related potentials (ERPs) that occur within milliseconds of error commission \cite{gehring1993neural, falkenstein1991effects}. These neural signatures have shown particular promise for human-computer interaction, with prior work demonstrating their potential for adaptive interface design \cite{zander2011towards, chavarriaga2010learning}. Previous work by Vi and Subramanian \cite{10.1145/2207676.2207744} demonstrated that error-related potentials can be reliably detected using consumer-grade EEG headsets in controlled settings, while Vi et al. \cite{10.1145/2556288.2557015} showed these signals can even be detected when observing others perform tasks. Eye-tracking studies have revealed characteristic patterns in gaze behaviour and pupil dilation associated with error awareness \cite{holmqvist2011eye, martinez2021application}, while heart rate variability measures derived from ECG have shown sensitivity to cognitive states that may predict error likelihood \cite{stuiver2012short, malik1996heart}.

\begin{figure*}[h]
    \centering
    \includegraphics[width=0.9\textwidth]{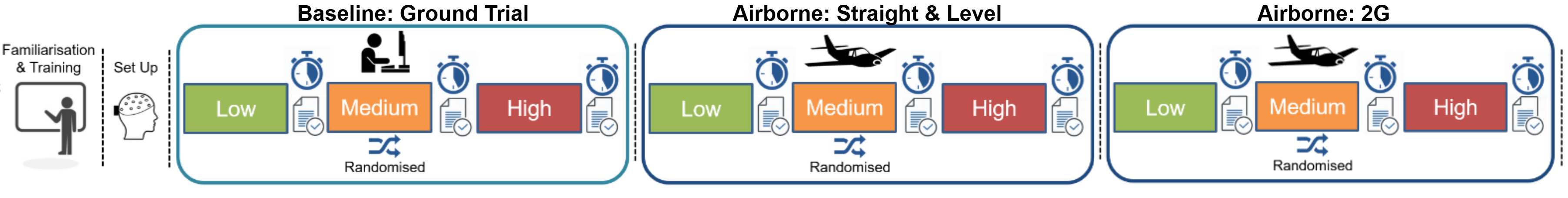}
    \caption{Overview of the Task Procedure}
    \Description{This figure outlines from left to right the process of the task procedure. Starting from the left is the initial setup and familiarisation with the task that each participant must go through. Once happy that they can complete the tasks they are then equipped with the neurophysiological sensors i.e. a wet EEG cap, a heart rate monitor, and eye-tracking glasses. The task then begins on the baseline i.e. ground trial, during this trial each of the low, medium, and high task settings are performed in a random order to limit mitigating factors. The process is then repeated for each of the airborne trials i.e. Straight \& Level as well as the 2G sustained spiral.}
    \label{experiment_potocol}
\end{figure*}

However, implementing physiological monitoring in operational environments presents significant challenges for human-computer interaction design. Aviation contexts introduce unique complications including motion artefacts, electromagnetic interference, and the need for unobtrusive sensor placement \cite{dehais2020neuroergonomics, wilson2005eeg}. Previous attempts to translate laboratory-based monitoring to real-world settings have highlighted the importance of robust signal processing and careful consideration of environmental factors \cite{kingphai2021eeg, han2020classification, mcguire2024deeper, McGuire2024PredictionOT}. This transition requires careful validation of both signal quality and classification reliability under actual operational conditions. Previous work in human-computer interaction has explored various approaches to error prevention in complex operational environments, from adaptive automation systems \cite{parasuraman2000model} to context-aware interfaces \cite{riley2016situation}. However, these approaches typically rely on behavioural or performance metrics rather than direct physiological indicators of error states. The integration of physiological monitoring could enable more proactive and personalised error prevention strategies \cite{sheridan2016human, hoc2000human}, but this requires understanding how well these measures perform in dynamic environments. Our work addresses this crucial gap by examining how physiological error detection methods translate to airborne environments, providing essential insights for the development of physiologically-adaptive interfaces in safety-critical systems.

\begin{figure}[]
    \centering
    \includegraphics[width=0.4\textwidth]{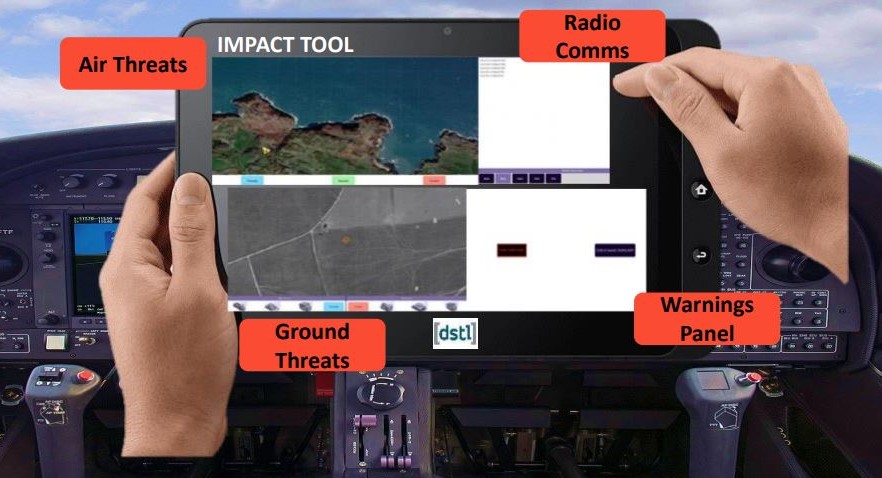}
    \caption{IMPACT Tool Visualisation}
    \Description{Figure \ref{IMPACT Tool Overview} IMPACT Tool Overview presents the IMPACT Tool Overview within a cockpit setting. The image shows a pilot's view of an aircraft cockpit, with hands holding a tablet device in the foreground. The tablet displays the IMPACT Tool interface, which is divided into several sections. At the top, there are areas labeled "Air Threats" and "Radio Comms". The central part of the screen shows a topographical map view labeled "IMPACT TOOL". Below this, there is a section for "Ground Threats" and a "Warnings Panel" on the right side. The tablet interface is superimposed over the cockpit view, which includes the aircraft's control yoke, instrument panels, and a view of a runway through the windshield.}
    \label{IMPACT Tool Overview}
\end{figure}

\section{Methodology}
\label{Study Overview}

\subsection{Participants}
To determine an appropriate sample size for our within-subjects study design, we conducted a power analysis considering medium effect size (f = 0.35), power of 0.80, and alpha level of 0.05, indicating a minimum requirement of eight participants (see Appendix Section \ref{Power Analysis}). To ensure we met all requirements for this investigation we recruited nine male commercial pilots through internal communications at BAE Systems. All participants held UK Commercial Pilot Licenses with DA42 aircraft clearance. Each assessment utilised a three-person crew: a participant performing cognitive tasks while wearing physiological sensors, an aircraft pilot performing manoeuvres and ensuring flight safety, and an equipment specialist managing data collection. Due to prescription eyewear requirements during flight, eye-tracking data was collected from four of the nine participants. The study protocol was approved by BAE Systems ethics review board.

\subsection{Environmental Scenarios}
The experimental design implemented a progressive series of testing environments to evaluate physiological error detection across increasingly challenging operational conditions. Beginning with a controlled laboratory baseline environment, we established fundamental error prediction capabilities without environmental stressors, providing critical reference data for validating physiological measurement reliability. This controlled setting served as the foundation for comparing subsequent airborne performance measures. The initial airborne testing phase utilised straight-and-level flight (S\&L), introducing participants to the basic challenges of flight operations including vibration, electromagnetic interference, and confined space considerations. During S\&L trials, a safety pilot maintained stable flight conditions while participants engaged in the multi-tasking scenario, enabling assessment of error detection capabilities under standard flight operations. This intermediate testing environment bridged the gap between laboratory conditions and more demanding flight scenarios. While environmental conditions necessarily progressed in order of increasing physical demands due to operational safety protocols, task difficulty levels (low, medium, high) were randomized within each environment to control for learning effects and time-dependent factors such as fatigue.

The most challenging testing environment involved sustained three-minute spiral manoeuvrers generating twice normal gravitational force (2G). These controlled spiral descents introduced substantial physical and vestibular challenges, enabling evaluation of error prediction robustness under extreme flight conditions \cite{han2020classification}. The 2G environment specifically tested the resilience of physiological measurement systems and error detection algorithms under conditions that significantly impact human cognitive and physiological functioning. All flight scenarios were conducted in a Diamond DA-42NG aircraft between 3,000 and 8,000 feet in non-pressurised conditions, ensuring consistent environmental parameters across trials. This systematic progression from controlled to increasingly complex environments facilitated a comprehensive evaluation of physiological error detection across realistic flight operations \cite{dehais2020neuroergonomics}. The unique combinations of motion effects, gravitational forces, and operational demands in each airborne environment provided an ideal context for validating these systems under conditions that closely approximate real-world aviation challenges.

\subsection{Task Design \& Data Collection}
To train and validate our error prediction models, we required accurately labelled timestamps of error occurrences aligned with subjects' physiological data. The IMPACT Tool \cite{sabine2022impact} (Figure \ref{IMPACT Tool Overview}) was selected over alternative multi-tasking paradigms for its distinct advantages in our flight research context. While established alternatives offer validated workload assessment, IMPACT provided superior flight-domain specificity with interfaces directly mirroring avionic systems encountered by our commercial pilot participants. Additionally, IMPACT's tablet deployment offered logistical advantages for cockpit integration compared to traditional desktop-based alternatives, particularly during challenging 2G maneuvers where space constraints and equipment security were critical concerns. The tool consists of four distinct interfaces, each designed to elicit and automatically log errors during specific sub-tasks.

The \textit{Radio Comms} interface challenges spatial cognition and auditory processing by requiring participants to track convoy positions through audio updates, recording errors when position classifications deviate from actual locations. The \textit{Ground Threats} section demands rapid visual discrimination of ground-based targets, measuring visual processing speed and classification accuracy with errors documented at misclassification or when targets move off-screen unclassified. 

\begin{figure}[]
    \centering
    \includegraphics[width=0.4\textwidth]{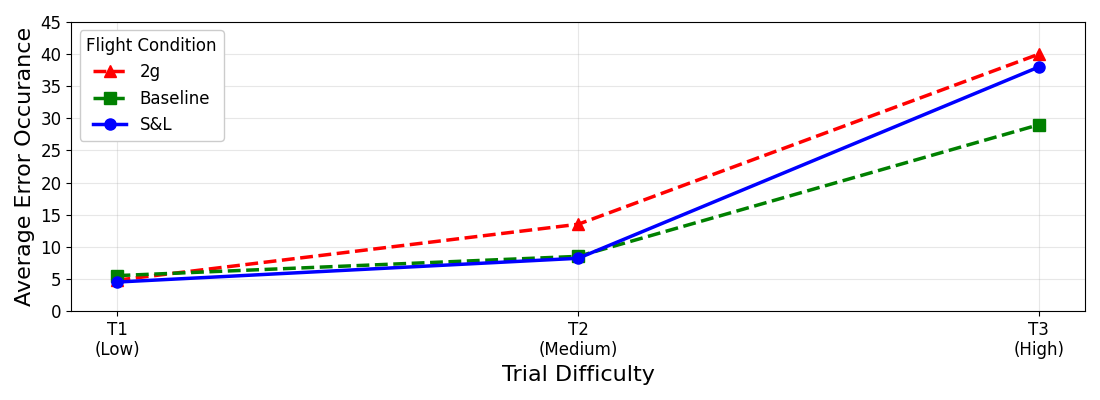}
    \caption{Average Occurrence of Errors across Trial Environments and Difficulty Levels Per-Participant}
    \Description{Figure \ref{experiment_potocol} A line graph showing the relationship between trial difficulty and average error occurrence across three different flight conditions. The x-axis represents trial difficulty with three levels: T1 (Low), T2 (Medium), and T3 (High). The y-axis shows Average Error Occurrence ranging from 0 to 45. Three lines are plotted, each representing a different flight condition: 2g (shown in red dashed line), Baseline (shown in green dashed line), and S&L (Straight & Level, shown in blue solid line). The graph demonstrates that error rates increase with trial difficulty across all flight conditions. At T1 (Low difficulty), all three conditions start with similar error rates around 5. At T2 (Medium difficulty), there is a modest increase in errors, with the 2g condition showing slightly higher rates (approximately 13) compared to Baseline and S&L (both around 8). The most dramatic increase occurs at T3 (High difficulty), where the 2g condition reaches around 40 errors, S&L reaches about 37 errors, and Baseline shows the lowest rate at approximately 29 errors.}
    \label{Average Occurrence of Errors}
\end{figure}

The \textit{Warnings Panel} focuses on sustained attention and response time through system monitoring, where participants must acknowledge alerts within a 5-second window, with errors marked at timeout events. Task difficulty was systematically manipulated across three levels (\textit{low, medium, high}) by adjusting event frequency and response windows to increase cognitive load and induce varying error rates, which it was successful in doing (see Figure \ref{Average Occurrence of Errors}), enabling evaluation of our models across different stress conditions. Each difficulty level was maintained for 3-minute intervals. The IMPACT Tool's logging system precisely timestamps all participant interactions, enabling accurate temporal alignment between error events and physiological responses. These timestamps serve as markers for extracting relevant windows of physiological data around error occurrences, facilitating the training of our machine learning models. To ensure balanced representation in our classification approach, error-free samples were uniformly selected from non-error events across all experimental conditions \cite{han2020classification}.

\subsection{Experimental Procedure}
Participants first received safety briefings and underwent physiological sensor fitting with three research-grade devices: a g.tec EEG system (24 channels, 256 Hz - sample rate) with wet gel-based electrodes, Tobii Pro 3 glasses for eye-tracking (100 Hz - sample rate), and a Polar H10 ECG chest strap (130 Hz - sample rate). All devices were selected based on CE certification and internal flight clearance requirements. Prior to data collection, participants completed multiple practice sessions with the IMPACT Tool until they reported confidence with all tasks and demonstrated consistent performance. The tablet-based IMPACT Tool was hosted on a Microsoft Surface Pro 7, securely mounted to the cockpit window using a suction cup holder for optimal viewing and interaction. Following familiarisation, trials progressed through three environments of increasing complexity. The baseline environment established reference measurements under controlled laboratory conditions without flight-related stressors. Subsequent straight-and-level (S\&L) flight introduced real aviation conditions while maintaining stable flight characteristics. Finally, the 2G spiral condition presented the highest physical and cognitive demands. Within each environment, task difficulty levels (low, medium, high) were randomised to control for learning effects, with each difficulty level maintained for 3-minute intervals. All flight operations were conducted in a Diamond DA-42NG aircraft between 3,000 and 8,000 feet in non-pressurised conditions, with continuous physiological data recording throughout all trials (see Figure \ref{experiment_potocol}).

\subsection{Physiological Data Processing}
Each physiological modality required preprocessing to address potential aviation-specific artefacts such as electromagnetic interference from cockpit instruments and motion artefacts from flight manoeuvres \cite{dehais2020neuroergonomics}. EEG data underwent bandpass filtering (0.5-50 Hz) to isolate relevant neural frequencies while removing muscle artefacts and electrical noise \cite{wilson2005eeg}, followed by re-referencing to improve signal quality for mobile EEG recordings \cite{kingphai2021eeg}. Eye-tracking data was processed using I-VT Fixation filtering, which has shown robust performance in dynamic environments for detecting saccades and fixations \cite{holmqvist2011eye}, while ECG data was filtered using established protocols for isolating cardiac components in noisy environments \cite{malik1996heart, hinde2021wearable}. For feature extraction, we employed 1-second sliding windows with no overlap, a duration shown effective for capturing error-related physiological responses \cite{chavarriaga2010learning} while maintaining temporal precision. For EEG, we extracted frequency bands (delta, theta, alpha, beta, gamma) and morphological characteristics shown to correlate with error detection \cite{gehring1993neural}. Eye-tracking features included fixation patterns, saccadic movements, and pupillometry metrics previously validated for cognitive state assessment \cite{beatty1982task}. ECG analysis focused on heart rate variability measures that reflect cognitive workload during error states \cite{stuiver2012short}. We explored both standard 1-second windows and extended 5-10 second windows with heart rate normalization for HRV metrics. Despite these methodological optimizations, ECG measures consistently performed near chance level (51.50\%) (See Table \ref{tab:model-performance}), suggesting cardiac signals may be too heavily influenced by the physical demands of flight to serve as reliable error indicators in this context (see Appendix \ref{Preprocessing} \& \ref{Feature Extraction} for complete details).

\subsection{Error Classification}
To assess if physiological data captured within the airborne environment serves as an effective marker of error occurrence, we implemented three machine learning approaches demonstrated effective with limited physiological datasets: Random Forest \cite{Pal2005RandomFC}, AdaBoost \cite{Hastie2009MulticlassA}, and Multi-layer Perceptron \cite{Popescu2009MultilayerPA, wirth2019towards}. These classifiers were selected based on three key considerations: their proven robustness with limited training data typical of specialized aviation studies, their established performance in prior physiological classification tasks \cite{wirth2019towards, mcguire2023song, mcguire2024prediction}, and their interpretability—a critical factor for initial validation in safety-critical contexts. Our focus was on establishing feasibility across varying environmental conditions rather than maximizing classification performance through more complex approaches. For all analyses, models classified 1-second windows (selected based on established physiological literature showing this duration optimally captures both error-related brain potentials occurring 50-500ms post-error \cite{falkenstein1991effects, gehring1993neural} and relevant eye movement patterns \cite{holmqvist2011eye} while maintaining feasible latency for cockpit applications \cite{zander2011towards}) as error or non-error events, using two balanced classes of data: timestamps of error events logged by the IMPACT Tool, and uniformly sampled timestamps of non-error-events. Each classifier was trained separately on individual physiological modalities (ECG, EEG, and Eye-Tracking). This study focused exclusively on binary error classification rather than exploring workload-error relationships, as our primary objective was to establish the fundamental feasibility of physiological error detection in airborne environments.

We conducted three complementary analyses to evaluate error detection performance: First, \textit{per-participant analysis} trained individual classifiers for each participant using their combined data from both ground and airborne trials, evaluated with 5-fold cross-validation. This approach assessed feasibility for personalized error detection by examining how well models perform for each specific individual (results shown in Figure \ref{EEG_Main}). Second, \textit{within-subject analysis} evaluated performance across all participants using 5-fold cross-validation while maintaining subject separation between folds \cite{kingphai2021eeg}. Unlike per-participant analysis, this approach trained a single model using data from all participants, testing how well a unified model could detect errors across different individuals. Third, \textit{between-subject analysis} employed leave-one-participant-out validation \cite{ancau2022deep}, training models on data from all participants except one, then testing on the held-out participant's data. This approach evaluated whether error detection systems could be deployed without individual calibration—critical for practical implementation in aviation. For all analyses, performance was evaluated using accuracy and F1-score, with statistical significance determined against a 50\% random baseline using t-tests (p = 0.001) \cite{Pagano2023BiasAU, zander2011towards}.

\section{Results}
\begin{table*}[h]
\caption{Average Model Performance Across Physiological Modalities Using Within-subject (CV: 5-fold Cross-validation) and Between-subject (LOO: Leave-one-participant-out) Validation Strategies}
\label{tab:model-performance}
\centering
\begin{tabular*}{\textwidth}{@{\extracolsep{\fill}}llccccc@{}}
\toprule
\textbf{Modality} & \textbf{Environment} & \textbf{Validation Strategy} & \textbf{Accuracy (\%)} & \textbf{Precision (\%)} & \textbf{Recall (\%)} & \textbf{F1-Score (\%)} \\
\midrule
\multirow{1}{*}{Random} & Both & Both & 50.00 $\pm$ 0.0 & 50.00 $\pm$ 0.0 & 50.00 $\pm$ 0.0 & 50.00 $\pm$ 0.00 \\
\midrule
\multirow{4}{*}{EEG} 
& Baseline & Within-subject (CV) & 89.23 $\pm$ 2.2* & 89.50 $\pm$ 1.7* & 86.60 $\pm$ 1.9* & 88.00 $\pm$ 1.00* \\
& Baseline & Between-subject (LOO) & 88.63 $\pm$ 2.3* & 88.20 $\pm$ 1.8* & 85.90 $\pm$ 2.0* & 87.00 $\pm$ 1.00* \\
& Airborne & Within-subject (CV) & 87.83 $\pm$ 2.4* & 87.70 $\pm$ 2.0* & 84.40 $\pm$ 2.1* & 86.00 $\pm$ 1.00* \\
& Airborne & Between-subject (LOO) & 86.80 $\pm$ 2.5* & 86.80 $\pm$ 2.2* & 83.30 $\pm$ 2.3* & 85.00 $\pm$ 2.00* \\
\midrule
\multirow{4}{*}{ET} 
& Baseline & Within-subject (CV) & 83.27 $\pm$ 3.1* & 83.50 $\pm$ 2.7* & 80.60 $\pm$ 2.8* & 82.00 $\pm$ 2.00* \\
& Baseline & Between-subject (LOO) & 82.47 $\pm$ 3.2* & 82.30 $\pm$ 2.8* & 79.80 $\pm$ 2.9* & 81.00 $\pm$ 2.00* \\
& Airborne & Within-subject (CV) & 82.50 $\pm$ 3.3* & 82.20 $\pm$ 2.9* & 79.90 $\pm$ 3.0* & 81.00 $\pm$ 2.00* \\
& Airborne & Between-subject (LOO) & 81.57 $\pm$ 3.4* & 81.40 $\pm$ 3.0* & 78.70 $\pm$ 3.1* & 80.00 $\pm$ 2.00* \\
\midrule
ECG & Both & Both & 52.22 $\pm$ 3.1 & 51.43 $\pm$ 3.7 & 51.98 $\pm$ 3.9 & 51.70 $\pm$ 3.01 \\
\bottomrule
\multicolumn{7}{l}{\small{* indicates statistical significance compared to random baseline (p < .001)}}
\end{tabular*}
\end{table*}

Our analysis revealed three key findings regarding the feasibility of physiological error detection in airborne environments: First, EEG demonstrated robust error detection capabilities across all flight conditions, achieving an average accuracy of 87.83\% ($\pm$2.4\%) during airborne trials using cross-validation, only slightly lower than baseline laboratory performance (89.23\% $\pm$2.2\%). The precision-recall balance remained strong in flight (87.70\% precision, 84.40\% recall), suggesting EEG signals maintain their diagnostic value despite additional noise and interference. This performance persisted even under leave-one-out validation (86.80\% accuracy, 86.80\% precision, 83.30\% recall for airborne), indicating potential generalisability across pilots.

Second, eye-tracking showed promising but more moderate performance, maintaining consistent accuracy around 82.50\% ($\pm$3.3\%) during flight with balanced precision (82.20\%) and recall (79.90\%). While lower than EEG, this reliability across conditions suggests eye movement patterns remain meaningful indicators of error states even during complex manoeuvres. However, the limited sample size (n=4) for eye-tracking warrants cautious interpretation.

Third, contrary to our expectations, ECG performed near chance level across all metrics (51.50\% accuracy, 51.30\% precision, 50.80\% recall), suggesting that cardiac measures may be too heavily influenced by the physical demands of flight to serve as reliable error indicators in this context. This finding highlights the importance of modality selection for operational implementations. Notably, performance remained stable across different flight conditions, with only minor degradation observed during 2G manoeuvres compared to straight-and-level flight (average decrease of 2.1\% for EEG, $p > .001$). This stability suggests that physiological error detection could potentially remain viable even during complex flight operations.

\begin{figure}[]
    \centering
    \includegraphics[width=0.5\textwidth]{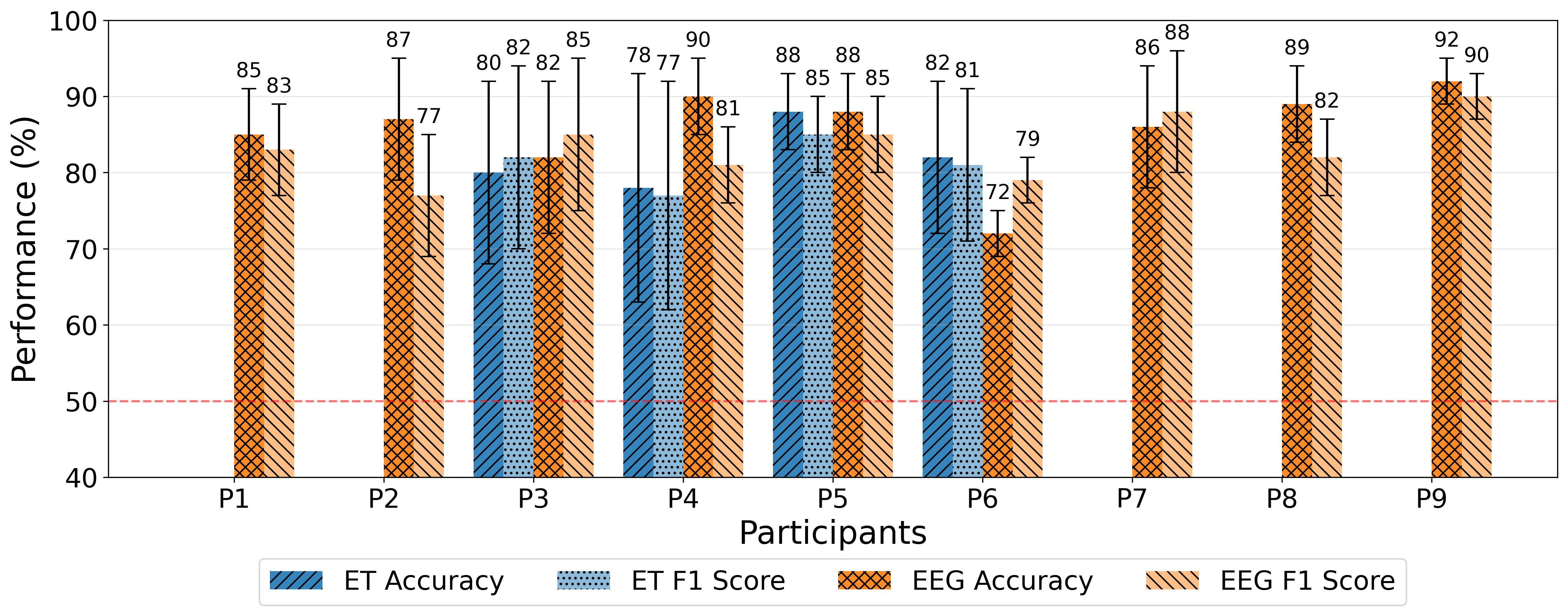}
    \caption{Average Performance Per-Participant for EEG and Eye-Tracking}
    \Description{Figure \ref{EEG_Main}  A bar graph comparing the performance metrics (Accuracy and F1 Score) between EEG and Eye-Tracking (ET) measurements across nine participants (P1-P9). The y-axis shows Performance percentage from 40\% to 100\%, with a red dashed line indicating the 50\% chance level. The x-axis lists participants P1 through P9. For each participant, up to four bars are shown: ET Accuracy (dark blue bars), ET F1 Score (light blue bars), EEG Accuracy (orange bars), EEG F1 Score (light orange bars) Not all participants have ET data, with P1, P2, P7, P8, and P9 showing only EEG measurements. For participants with both measurements (P3-P6), the performance values generally range between 70\% and 90\%. Error bars are included for all measurements, showing the variance in performance. All participants performed well above the 50\% chance level, with most scores falling between 75\% and 90\%. The highest performances were seen in P9, with EEG accuracy reaching 92\%.}
    \label{EEG_Main}
\end{figure}

Analysis of individual participant performance revealed consistent but varying levels of success across both EEG and eye-tracking modalities (Figure \ref{EEG_Main}). For EEG, classification accuracy ranged from 72\% to 92\% across participants, with most participants (7 out of 9) achieving accuracy above 85\%. Notably, participant P9 achieved the highest EEG performance (92\% accuracy, 90\% F1-score), while P6 showed the lowest (72\% accuracy, 79\% F1-score), suggesting some individual variability in signal quality or task engagement. Eye-tracking data, available for participants P3-P6, showed comparable though slightly lower performance levels, with accuracies ranging from 78\% to 88\%. The strong correlation between accuracy and F1-scores (r = 0.91, p < .001) across both modalities suggests robust classification performance regardless of class distribution. Interestingly, participants with both EEG and eye-tracking data (P3-P6) showed complementary patterns – when EEG performance decreased, eye-tracking often maintained higher accuracy, suggesting potential benefits of multi-modal approaches. All participants significantly exceeded chance-level performance (50\%, indicated by red dashed line) across both modalities (p < .001), demonstrating the robustness of physiological error detection even in challenging flight conditions. The consistency of F1-scores with accuracy metrics indicates balanced performance across error and non-error states, an important consideration for operational implementation.

\section{Discussion \& Conclusion}
Our findings demonstrate the feasibility of using physiological signals to detect operator errors in airborne environments, with significant implications for human-computer interaction in aviation contexts. The robust performance of EEG-based classification (87.83\% accuracy) during airborne testing suggests promising opportunities for real-time error detection systems \cite{dehais2020neuroergonomics}. The success of both EEG and eye-tracking measures opens possibilities for context-aware cockpit interfaces that could adapt to pilot cognitive states before errors manifest in behaviour \cite{parasuraman2000model}. Three key implications emerge from our analysis. First, the high accuracy of EEG-based detection in airborne environments demonstrates that laboratory-derived approaches can translate effectively to dynamic settings \cite{wilson2005eeg}. This validates the potential for physiologically-adaptive interfaces in operational contexts. Second, the complementary performance of EEG and eye-tracking suggests value in multi-modal approaches, where different measures could provide redundant error detection capabilities \cite{han2020classification}. Third, the poor performance of ECG measures (51.50\%) highlights crucial challenges in separating cognitive from physical responses during flight operations, emphasising the importance of careful sensor selection for operational implementations \cite{harris2006meta}.

Several limitations of our current study warrant discussion. A methodological limitation was our use of a fixed progression through environmental conditions rather than a counterbalanced design, necessitated by flight safety protocols and operational constraints. This fixed ordering could introduce time-dependent effects where vigilance decrements, fatigue, and learning might influence physiological measures—particularly impacting EEG theta and alpha bands \cite{gevins1997high}, pupillometry responses \cite{beatty1982task}, and ECG metrics \cite{stuiver2012short}. While we mitigated these effects by randomizing difficulty levels within each environment, future research should consider counterbalanced designs where operational safety permits. Our participant sample (n=9) represents a highly specialised population of commercial pilots with DA42 aircraft clearance, with recruitment constrained by strict aviation safety protocols and the high operational costs of live flight trials. The limited eye-tracking dataset (n=4) resulted from prescription eyewear requirements—a common challenge in operational aviation contexts where safety equipment takes precedence over research instrumentation. While these sample sizes are modest, they are consistent with similar in-flight research protocols \cite{dehais2017pilot, wilson2005eeg}, and our power analysis indicated sufficient statistical strength for our primary measures. Additionally, our use of standardised tasks rather than actual flight operations potentially limits ecological validity, while our focus on post-error classification rather than pre-emptive error prediction limits applications for real-time intervention. Future work should investigate how both our current approach and potential pre-emptive systems could be integrated into operational cockpits, addressing challenges including non-intrusive sensor placement, appropriate intervention thresholds, and pilot-centred feedback mechanisms \cite{dehais2017pilot}. Critical to successful implementation will be understanding how pilots interact with and trust such systems during actual flights \cite{parasuraman2000model}. This research direction could fundamentally advance aviation safety by detecting physiological precursors before errors manifest—a capability our work demonstrates is feasible within airborne environments.

In conclusion, this work provides the first demonstration that physiological error detection can maintain high accuracy during actual flight operations, offering new possibilities for human-computer interaction in safety-critical environments. Our findings that EEG-based classification achieves 87.83\% accuracy even during complex manoeuvrers, while eye-tracking maintains 82.50\% accuracy, suggest that laboratory-derived approaches can successfully translate to dynamic operational contexts. The complementary performance patterns between modalities, coupled with insights about the limitations of ECG measures (51.50\%), provide crucial guidance for developing robust error detection systems in real-world settings. While challenges remain in translating these findings to operational systems, our work establishes a foundation for future research into adaptive interfaces that could enhance aviation safety through early error detection and intervention.

\begin{acks}
The authors would like to express their gratitude to BAE System Air Sector Human Factors and Technology Research Team for their invaluable support and collaboration in this research. In particular, we extend our sincere appreciation to Dr Jacob Greene for his insightful contributions, technical expertise, and guidance throughout the development of this work. His input has been instrumental in refining our approach and ensuring the practical relevance of our findings. This work was supported by the Engineering and Physical Sciences Research Council [grant number EP/W522260/1]. Results were obtained using the ARCHIE-WeSt High Performance Computer\footnote{www.archie-west.ac.uk} based at the University of Strathclyde.

\end{acks}

\bibliographystyle{ACM-Reference-Format}
\bibliography{ref}

\appendix

\section{Power Analysis}
To determine the required sample size for our within-subjects study design, we conducted an a priori power analysis using logistic regression parameters. With an odds ratio of 1.5, \(\alpha \) = 0.05, desired power (1-\(\beta \)) = 0.80, and assuming a normal distribution of predictors, the analysis indicated a minimum requirement of 8 participants to achieve adequate statistical power. Our final sample of 9 participants exceeded this minimum threshold, ensuring sufficient power for our analyses.

\label{Power Analysis}
\begin{figure}[H]
    \centering
    \includegraphics[width=0.7\linewidth]{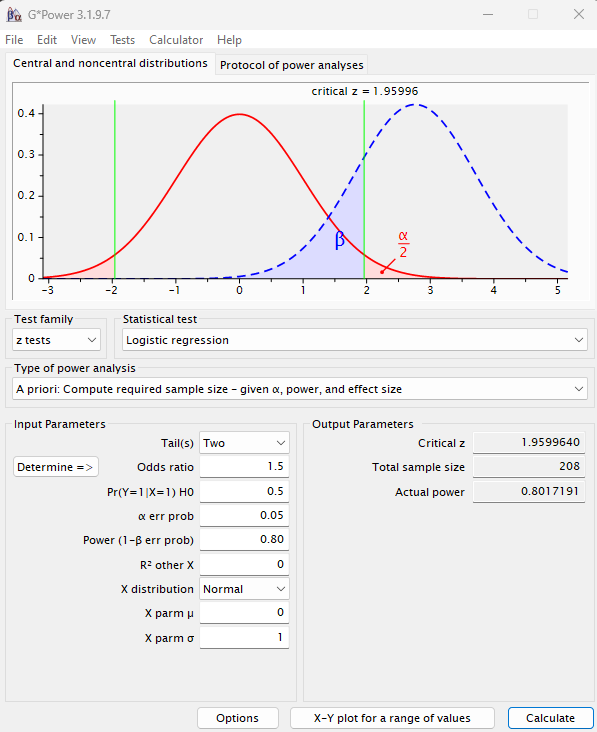}
    \caption{Power Analysis of Required Number of Participants}
    \label{fig:enter-label}
\end{figure}

\section{Physiological Data Preprocessing}
\label{Preprocessing}
Raw physiological signals often contain artefacts and noise from motor functions, electrical interference, and environmental factors \cite{kingphai2021eeg, luck2014introduction}.
\subsection{EEG}
\label{sec: EEG_preprocessing}
Our autonomous preprocessing methods \cite{kingphai2021eeg, wilson2005eeg, herrmann2016eeg} include sequential filtering implemented through MNE-Python: bandpass filtering (0.5-50 Hz using FFT-based FIR filter, transition bandwidth 0.1 Hz and 0.5 Hz for low and high respectively), and average re-referencing across all scalp channels \cite{yao2001reference, chella2016calibration}. Baseline correction was applied using whole-signal mean subtraction. Data was processed using 4 parallel jobs for computational efficiency, with signals sampled at 256 Hz.
\subsection{ECG}
\label{sec: ECG_preprocessing}
ECG preprocessing employs bandpass filtering (0.5-40 Hz, FFT-based FIR filter) to isolate signals while preserving cardiac components, and Discrete Wavelet Transform (DWT) for noise reduction. Signal processing was performed on 130 Hz sampled data using custom Python implementations of the firwin filter design.
\subsection{ET}
\label{sec: ET_preprcessing}
Eye-tracking data preprocessing uses Tobii Pro Lab with I-VT Fixation filter (velocity threshold: 30°/s). For gaze patterns, we apply median filtering (kernel size 3, 100ms window) for smoothing, constant interpolation for gaps <75ms, and three-point velocity computation over 20ms windows. Pupillometry processing involves initial smoothing with a rolling median filter (3-sample window), artefact detection for rapid changes (>1 mm/s), secondary smoothing (5-sample median filter), and cubic spline interpolation for gaps <200ms. Data was processed at the native 100 Hz sampling rate.

\section{Feature Extraction}
\label{Feature Extraction}
To address computational challenges with high-dimensional physiological data \cite{han2020classification}, we extracted key features from EEG, ET, and ECG data as detailed below.

\subsection{EEG Features}
\label{sec: Feature Extraction_EEG}
EEG features were extracted using a 1-second sliding window with 256 Hz sampling (512 samples per window) \cite{wirth2019towards}. We computed six feature categories: (1) Frequency domain features including PSD bands (delta: 1--4 Hz, theta: 4--8 Hz, alpha: 8--12 Hz, beta: 12--30 Hz, gamma: 30--50 Hz) \cite{klimesch1999eeg}; (2) Statistical features (mean, variance, skewness, kurtosis) \cite{teplan2002fundamentals}; (3) Morphological features capturing signal shape through curve length, peak count, and non-linear energy across 1--50 Hz \cite{gevins1997high}; (4) Time-frequency features from wavelet transforms \cite{herrmann2016eeg}; (5) Linear features using autoregressive coefficients (AR, p = 2) \cite{wilson2005eeg}; and (6) Non-linear features including approximate entropy and Hurst exponent \cite{kingphai2021eeg}.

\subsection{Eye Tracking Features}
\label{sec: Feature Extraction_ET}
Using Tobii eye-tracking glasses \cite{holmqvist2011eye}, we extracted two main feature types. First, gaze pattern features captured saccadic movements (identified by velocity threshold >25°/s) \cite{duchowski2007eye, martinez2021application} and fixation points \cite{majaranta2014eye}. Second, blink features were detected through median-filtered pupillometry data within 70--450ms windows \cite{beatty1982task, wilson2002cardiac}.

\subsection{ECG Features}
\label{sec: Feature Extraction_ECG}
ECG feature extraction focused on two key aspects \cite{malik1996heart, hinde2021wearable}. First, statistical features characterised the ECG signal distribution through mean, standard deviation, skewness, and kurtosis \cite{stuiver2012short, qu2021classification}. Second, RR interval analysis provided heart rate variability metrics including mean interval duration, SDNN (standard deviation of intervals), and coefficient of variation \cite{malik1996heart, hinde2021wearable}.

\end{document}